\documentclass
[prb,showpacs,byrevtex,twocolumn,tightenlines,10pt,a4paper,groupedaddress,bibnotes,footinbib,preprint]{revtex4}%
\usepackage{amsfonts}
\usepackage{amsmath}
\usepackage{amssymb}
\usepackage{graphicx}%
\begin{document}
\title{Conductance noise in interacting Anderson insulators driven far from
equilibrium }
\author{V. Orlyanchik, and Z. Ovadyahu\vspace{0.001in}}
\affiliation{Racah Institute of Physics, The Hebrew University, Jerusalem 91904,
Israel\vspace{0.002in}\vspace{0.002in}}

\begin{abstract}
The combination of strong disorder and many-body interactions in Anderson
insulators lead to a variety of intriguing non-equilibrium transport
phenomena. These include slow relaxation and a variety of memory effects
characteristic of glasses. Here we show that when such systems are driven with
sufficiently high current, and in liquid helium bath, a peculiar type of
conductance noise can be observed. This noise appears in the conductance
versus time traces as downward-going spikes. The characteristic features of
the spikes (such as typical width) and the threshold current at which they
appear are controlled by the sample parameters. We show that this phenomenon
is peculiar to hopping transport and does not exist in the diffusive regime.
Observation of conductance spikes hinges also on the sample being in direct
contact with the normal phase of liquid helium; when this is not the case, the
noise exhibits the usual 1/f characteristics independent of the current drive.
A model based on the percolative nature of hopping conductance explains why
the onset of the effect is controlled by current density. It also predicts the
dependence on disorder as confirmed by our experiments. To account for the
role of the bath, the hopping transport model is augmented by a heuristic
assumption involving nucleation of cavities in the liquid helium in which the
sample is immersed. The suggested scenario is analogous to the way high-energy
particles are detected in a Glaser's bubble chamber.%
\vspace*{\fill}%

\end{abstract}
\maketitle

\section{Introduction}

Conduction noise is an inherent property of essentially all electronic
systems. The most common form of this noise has a 1/f$^{\alpha}$
power-spectrum (PS) with $\alpha$ of order unity \cite{1}. The ubiquity of the
1/f spectrum in the noise of Fermi-gas systems is quite intriguing in that it
seems to be insensitive to the specific type of transport. For example, as the
metal insulator transition is crossed, the transport mode changes from
diffusive to a hopping process, and this is preceded by a dramatic increase of
the noise magnitude \cite{2}. However, the power spectrum usually retains its
power-law form with $\alpha$ changing only slightly \cite{2,3}.

In this paper, we describe results of noise experiments performed on Anderson
localized indium-oxide films measured at liquid helium temperatures, deep in
the hopping regime. Several versions of indium-oxide films were employed in
these studies; crystalline samples as well as several variants of amorphous
indium-oxide (\textit{e.g}., different carrier concentrations). Electronic
transport in these systems was extensively studied in the near-equilibrium
regime as well as when driven far from equilibrium \cite{4}. In the latter
case, peculiar glassy features were found in accordance with theoretical
expectations for strongly interacting Anderson insulators \cite{5}. In
particular, when excited far from equilibrium, the conductance of the system
increases, and then relaxes slowly towards its equilibrium value. This
relaxation time is controlled by several factors the most important of which
is the strength of the inter-electron interaction - the stronger the
interaction the more sluggish is the relaxation \cite{6}. Increasing the
static disorder (characterized, say, by the sample resistivity at a given
temperature) also slows down the relaxation and so does a high magnetic field
\cite{7}. In addition, other non-equilibrium features were observed such as
aging and related memory effects \cite{8}. The original motivation for
studying conductance noise in these systems was to get more information on
their glassy behavior.

The above glassy features are restricted to the strongly localized regime and
they disappear when the system crosses over to the diffusive regime by either,
reducing the disorder or raising the sample temperature \cite{9}. In this
crossover from ergodic transport to a glassy transport regime, the noise
characteristics of these samples are essentially unaffected. In particular,
the noise is Gaussian and has a 1/f$^{\alpha}$ power spectrum with $\alpha$
close to 1 on \textit{both} sides of the `ergodic-to-glassy' crossover, again
illustrating the ubiquity of this type of noise. In addition, no
\textquotedblleft saturation\textquotedblright\ is observed in the PS down to
f=10$^{-3}$ Hz as shown in figure 1 contrary to theoretical expectations that
assumes that the noise is due to fluctuations in carrier concentration
\cite{10}.%

\begin{figure}
[ptb]
\begin{center}
\includegraphics[
trim=0.000000in 1.290083in 0.000000in 0.581507in,
natheight=6.259500in,
natwidth=4.371600in,
height=3.4368in,
width=3.4247in
]%
{wmf/Graph1.WMF}%
\caption{Typical Power Spectrum of conductance fluctuations in an insulating
sample, measured under linear response condition at $T$=4.11K. Sample:
InO$_{x}$, length=1mm, width=1mm, $R_{_{\square}}$=12M$\Omega$. The lower
frequency part of the spectrum was measured by averaging 36 1-hour pieces of
$G(t)$ runs and that was spliced with two spectra taken with the HP35660A up
to 800Hz. The dashed line represents a 1/f power-spectrum.\bigskip}%
\end{center}
\end{figure}

It is important to emphasize however that the 1/f spectrum such as shown in
figure 1 is observed when the noise is measured on \textit{macroscopic}
samples \cite{11} and under \textit{linear-response} conditions. On the other
hand, when the drive current through the sample exceeds a threshold value, and
in addition, the sample is immersed in the normal phase of liquid helium, the
noise characteristics change dramatically.\newline%
\begin{figure}
[ptb]
\begin{center}
\includegraphics[
trim=0.000000in 1.289278in 0.000000in 0.580300in,
natheight=6.246500in,
natwidth=4.356100in,
height=3.435in,
width=3.4195in
]%
{wmf/Graph2.WMF}%
\caption{Conductance noise Power Spectra at different bias currents. Note the
abrupt change of the noise characteristics between $I_{T}$=1.9$\mu$A and
$I_{T}$=4.4$\mu$A. The dashed line is a 1/f PS for comparison. Sample:
In$_{2}$O$_{3-x}$, length=3.5mm, width=1mm, $R_{_{\square}}$=20M$\Omega
$.\bigskip}%
\end{center}
\end{figure}

As shown in figure 2 the PS becomes flat (\textquotedblleft
saturated\textquotedblright) up to some corner frequency f* and above it the
PS drops sharply, faster than a power-law. The transition is also
characterized by a considerable increases in the noise magnitude, especially
around f*. Time domain sweeps (figure 3) reveal that this excess noise is
associated with the appearance of downward-going spikes in the conductance
whose (average) frequency increases exponentially with the drive current.%
\begin{figure}
[ptb]
\begin{center}
\includegraphics[
trim=0.000000in 1.289278in 0.000000in 0.580300in,
natheight=6.246500in,
natwidth=4.356100in,
height=3.435in,
width=3.4195in
]%
{wmf/Graph3.WMF}%
\caption{Conductance as a function of time in ac (using a sine drive at
\ 23Hz, upper plot) and dc (lower plot) measurement configurations. Note the
downward-going spikes in both cases. Sample: InO$_{x}$, length=1mm, width=1mm,
$R_{_{\square}}$=12.5M$\Omega$.\bigskip}%
\end{center}
\end{figure}

The phenomenology associated with this new type of noise is described in
section III below. It is demonstrated that such current spikes may be
reproduced by artificially generated cooling-bursts. We then present a
heuristic picture that explains how \textit{spontaneous} cooling-events may
arise from the interplay between the sample being driven far from linear
response, and its interaction with the liquid helium bath. Specifically, the
cooling events are ascribed to the production of cavities at the sample/liquid
interface. These are triggered by high-energy events that intermittently
appear in the current-driven sample. It is shown that the statistical
occurrence of such events is a natural consequence of hopping transport in a
strongly interacting system. A model, based on the percolation picture of
hopping transport, shows how such events arise and predicts their occurrence
probability as function of current and sample parameters. This is detailed in
section IV. The model considers the charge transport in the hopping system as
a traffic-flow in a network that, upon strong enough drive, results in
traffic-jam events in analogy with other physical situations. Flow of
particles through a disordered system often leads to traffic congestion
problems resulting from the interplay between disorder and interactions
\cite{12}. The most familiar example for this phenomenon is traffic-jams that
are part of modern life in urban areas. A common form of this problem occurs
when the density of cars in a one-lane road exceeds a threshold value. As a
rule, this will result in a `stop-and-go' traffic flow although a continuous
motion at the average speed of the slowest driver is theoretically possible.
Granular flow is another example of a similar nature, which due to its
technological implications has received wide attention \cite{13,14}. Here, we
present experimental results that are consistent with traffic-jam behavior in
disordered electronic systems. It is argued that this phenomenon is generic to
Anderson insulators where transport is by hopping and where Coulomb
interaction is significant. These are the same two ingredients that lead to
electron-glass behavior. Note however that the effects associated with
electron glass are measured in the linear response regime, and they actually
disappear \cite{15} when driven too far into the non-ohmic regime. In
addition, none of the glassy effects requires the presence of liquid helium
for their observation. The feature that is common to the two phenomena is that
hopping conductivity is a pre-requirement as both phenomena disappear in the
diffusive regime.

In section V we describe a heuristic picture that purports to explain the way
gaseous cavities at the sample/helium interface are generated in response to
the electronic jams, and further results and discussion is given in section VI.

\section{II - Samples description and measurements techniques}

Our experiments were performed using either crystalline or amorphous
indium-oxide films (referred to in this paper as In$_{2}$O$_{3-x}$ and
InO$_{x}$ respectively). These were prepared and their disorder fine-tuned to
be in the insulating regime by the methods described elsewhere \cite{16}. Most
of the samples were deposited on microscope glass-slides. However, identical
results were obtained using samples deposited on alumina or Si-wafers as
substrates. The samples thickness was typically 50$\mathring{A}$, and
200$\mathring{A}$, for In$_{2}$O$_{3-x}$ and InO$_{x}$ respectively, and
various lateral dimensions of 20$\mu$m to 3cm were studied. Except when
otherwise noted, all measurements reported here were done with the sample
immersed in liquid helium at $T$=4.11K (under 690 Torr). The samples were
mounted on a stage at the end of a probe and were loosely wrapped with a
Teflon tape. This was done for protecting the sample and for securing the
connecting wires. However, we found that the Teflon tape also contributed to
the consistency of the spikes appearance, probably by protecting the sample
from spurious thermal shocks.

Conductance measurements were made by biasing the sample with a constant
voltage source while measuring the resulting current (voltage drop across a
series resistor). In addition to this two-terminal configuration, we employed
in some cases a four-terminal technique to verify that the measured noise
originates from the sample (rather than from the contacts). Time traces and
spectra were recorded by HP35660A or HP35670A Spectrum Analyzers using EG\&G
5113 as pre-amplifier.

In most of the experiments the helium bath was a storage dewar, which was
convenient for long measurements at a constant stable temperature. Based on
previous studies using these materials we know that insulating In$_{2}%
$O$_{3-x}$ films exhibit the Mott's VRH law: $R(T)\propto\exp\left(
\frac{T_{0}}{T}\right)  ^{\frac{1}{3}}$. For samples with $R_{_{\square}}$
between 10M$\Omega$ and 10G$\Omega$ at 4K, which is the range studied here,
$T_{0}$ was in the range of 3000-8000K. The hopping law in the amorphous films
depends, among other things, on the carrier concentration $n$. Samples with
$n>10^{20}cm^{-3}$ (a range of $n$ which exhibited spikes most clearly), tend
to exhibit $R(T)\propto\exp\left(  \frac{T^{\ast}}{T}\right)  ^{\frac{1}{2}}$
with $T^{\ast}$ ranging between 150 to 600K for the above range in
$R_{_{\square}}$.

\section{III - The basic features of the phenomenon}

The phenomenon we wish to focus can be observed in the time dependence of the
sample conductance. This is illustrated in figures 4 and 5 for typical
In$_{2}$O$_{3-x}$ and InO$_{x}$ samples respectively. For sufficiently small
drive current $I_{T}$, the conductance $G$ versus time shows only small
fluctuations, which, upon Fourier transforming turn out to exhibit the common
1/f noise power-spectrum \cite{2}. However, once $I_{T}$ exceeds a threshold
value, a sudden change in the time traces is observed as illustrated in
figures 4 and 5.%
\begin{figure}
[ptb]
\begin{center}
\includegraphics[
trim=0.000000in 0.771443in 0.000000in 0.129302in,
natheight=6.246500in,
natwidth=4.356100in,
height=4.1874in,
width=3.4195in
]%
{wmf/Graph4.WMF}%
\caption{Conductance fluctuations as a function of time for a typical In$_{2}%
$O$_{3-x}$ sample measured at different bias currents (the current values in
$\mu$A are shown for each trace) near the threshold. Sample length=1.1mm,
width=3mm, $R_{_{\square}}$=68M$\Omega$ (at the threshold drive).}%
\end{center}
\end{figure}
\begin{figure}
[ptbptb]
\begin{center}
\includegraphics[
trim=0.000000in 1.159350in 0.000000in 0.389157in,
natheight=6.246500in,
natwidth=4.356100in,
height=3.6867in,
width=3.4195in
]%
{wmf/Graph5.WMF}%
\caption{Conductance fluctuations as a function of time for a typical
InO$_{x}$ sample measured at different bias currents (the current values in
$\mu$A are shown for each trace) near the threshold. Sample: length=0.4mm,
width=2mm, thickness=175$\mathring{A}$, $R_{_{\square}}$=75M$\Omega$ (at the
threshold drive).\bigskip}%
\end{center}
\end{figure}

Note that above a certain $I_{T}$ the conductance versus time traces $G(t)$
acquire an additional component in the form of downward-going spikes that are
rare near the threshold $I_{T}$ but their number per unit time increases
dramatically with current. A mere 3-4\% increase in $I_{T}$ past the threshold
typically increases the spikes frequency by two orders of magnitude
(\textit{c.f.}, figures 4\&5 and the discussion regarding this issue in
section V).

Spikes do not appear (in the experimental time-window of up to 100sec.) unless
the sample is driven with sufficiently high current. As a rule, this occurs
when the sample is far into the non-ohmic regime as shown in figure 6. The
first observation is then that this is manifestly a far-from-linear response
phenomenon.%
\begin{figure}
[ptb]
\begin{center}
\includegraphics[
trim=0.000000in 1.420907in 0.000000in 0.582133in,
natheight=6.259500in,
natwidth=4.371600in,
height=3.3347in,
width=3.4247in
]%
{wmf/Graph6.WMF}%
\caption{Resistance as a function of the bias field for amorphous (InO$_{x}%
$-solid circles, length=0.5mm, width=5mm) and crystalline (In$_{2}$O$_{3-x}%
$-open squares, length=0.15mm, width=1mm) samples. The arrows mark the
threshold for the appearance of spikes.\bigskip}%
\end{center}
\end{figure}

In the majority of samples, (more than 60 studied here) the spikes had the
same generic shape; a fast \textquotedblleft attack\textquotedblright%
\ followed by a longer \textquotedblleft recovery-tail\textquotedblright. The
relative magnitude of the spike and its duration however, are dependent on
specific sample parameters as discussed later. Naturally, the appearance of
these spikes modifies the noise characteristics of the signal in terms of
both, magnitude and spectrum: The power spectrum turns out to be flat
(\textquotedblleft white-noise\textquotedblright) at low frequencies, followed
by a fast decline above a certain frequency, both features are obviously
related to the peculiar shape of the individual spike. In many cases, the
spikes had the same unique shape (as, e.g., in figure 4) namely; each spike is
an exact replica of any other spike. In most other instances, two or three
different shapes could be identified. The detailed shape of a particular spike
is sample specific; Samples from the same batch with nearly identical
$R_{_{\square}}$ and $I-V$ characteristics can be distinguished by their spike shape.

These remarks apply to the immediate vicinity of the threshold where
individual events could be resolved in the time domain, which was an extremely
narrow range of bias current \cite{17}. At higher currents, the power spectrum
change with $I_{T}$ indicates that the spikes duration may become shorter
(\textit{c.f.}, the shift of f* to higher frequencies in figure 2).

Figure 7 shows time traces at different currents employing a 4-terminal
measurement on an insulating In$_{2}$O$_{3-x}$ sample. The spikes in this case
exhibit the same shape as those in figures 4 and 5, and their number per unit
time increase with the drive current in the same manner as before. However,
these spikes are now \textit{upward}-going. One may therefore conclude that
the events associated with the spikes are temporary drops of the sample
conductance occurring in the bulk of the film (as opposed to contacts
problems).%
\begin{figure}
[ptb]
\begin{center}
\includegraphics[
trim=0.000000in 1.407763in 0.000000in 0.577369in,
natheight=6.201600in,
natwidth=4.291200in,
height=3.333in,
width=3.3918in
]%
{wmf/Graph7.WMF}%
\caption{Four probe measurement of resistance as a function of time below and
above the threshold current (the current values in $\mu$A are shown for each
trace). Sample: In$_{2}$O$_{3-x}$, length=0.15mm, width=15mm, with
$R_{_{\square}}$=3.7M$\Omega$ (at the threshold drive).\bigskip}%
\end{center}
\end{figure}

Similar spikes and associated power spectrum had been observed in flow of
granular particles through a pipe \cite{18}. We believe that this similarity
is not coincidental - both phenomena involve an intermittent dropout in
traffic-throughput due to a forced-flow of strongly interacting, discrete
particles. However, the physical processes that lead to the observation of
conductance spikes are more intricate than the one-dimensional granular flow
problem. The complex nature of the this phenomenon may be appreciated by
considering the following key observations:

1.\qquad The bath in which the sample is immersed plays a crucial role. Spikes
do not appear unless the sample is in contact with liquid helium. Also, no
spikes are observed when the sample is vacuum-loaded (while being kept at
$T$=4.11K via a copper cold-finger). In addition, coating the sample with a
thick layer of a photo-resist film eliminated the spikes even when the sample
was immersed in liquid helium. A more dramatic demonstration of the role of
the bath is illustrated in figure 8. This shows that the spikes vanish
abruptly upon cooling the helium bath below $T$=2.174K. Note that this
temperature is the `$\lambda$-point' below which the helium becomes a
superfluid.%
\begin{figure}
[ptb]
\begin{center}
\includegraphics[
trim=0.000000in 1.408383in 0.000000in 0.576749in,
natheight=6.201600in,
natwidth=4.291200in,
height=3.333in,
width=3.3918in
]%
{wmf/Graph8.WMF}%
\caption{$\delta G(t)$ for different temperatures near the $\lambda$-point.
Note the abrupt disappearance of the spikes at the $\lambda$-point. Sample:
In$_{2}$O$_{3-x}$, length=0.5mm, width=0.5mm, $R_{_{\square}}$=16M$\Omega$ (at
the threshold drive).}%
\end{center}
\end{figure}

2.\qquad The phenomenon is peculiar to samples that are deep into the hopping
regime having sheet resistance of 1M$\Omega$ to 10G$\Omega$. No spikes
appeared in any of our In$_{2}$O$_{3-x}$ samples that were in the diffusive
regime (\textit{i.e.}, when the sample conductance $G$ is higher than the
quantum-value $e^{2}/h$) even at currents that were so high as to cause helium
boiling (six orders of magnitude higher power than that used in figures 4 or
5). Therefore, the mode of transport in the sample plays an important role in
giving rise to the conductance spikes.

These empirical observations suggest that the spikes involve interplay between
the sample and the liquid bath, and in the following, we attempt to elucidate
their respective role.

The disappearance of the spikes at the transition into the superfluid phase
led us to suspect that the spikes may be associated with thermal events, which
are then annihilated (or considerably weakened) by the superfluid component
kicking-in as the temperature falls below the $\lambda$-point. To account for
the observed downward-going spikes, these have to be \textit{cooling}-events
given the fact that the conductance in the hopping regime increases with
temperature (namely, $dG/dT>0$). The fractional change of the conductance
$\Delta G/G$ associated with a spike is 0.01-0.2\% which could be affected by
a cooling-burst of 0.1-2mK for a typical sample. To check on such a `cooling
conjecture', several samples were subjected to cooling-bursts produced by
cutting-off the power, for brief period ($\approxeq$2.5mS), from an external
heat source to which the sample was otherwise constantly exposed. Typical
results are illustrated in figure 9 using two different ways to create cooling
events. Adjusting the duty-cycle such that power into the heat source is
applied only during 2.5mS intervals resulted instead in upward-going spikes
(figure 9). The heat source in these examples was either a micro-resistor or
small IR light-emitting diode coupled thermally to the sample. The
micro-resistor was a 350$\mathring{A}$~gold film 20$\mu$m wide and 10mm long
deposited on a mylar film which was thermally anchored to the sample.%
\begin{figure}
[ptb]
\begin{center}
\includegraphics[
trim=0.000000in 1.417331in 0.000000in 0.542196in,
natheight=6.246500in,
natwidth=4.356100in,
height=3.3658in,
width=3.4195in
]%
{wmf/Graph9.WMF}%
\caption{Spikes caused by artificial cooling and heating events: (a). IR LED
cooling; (b). IR LED heating; (c). Micro-resistor heating; (d). Micro-resistor
cooling. The long conductance tail observed after the IR excitation in (b) is
an inherent glassy effect; This slow relaxation occurs whenever the sample is
excited with sufficiently high quantum energy source (\textit{c.f.},
Ben-Chorin et al in reference 6). Note the absence of such tail in plate c.
Samples: (a) and (b) InO$_{x}$, length=0.6mm, width=1.5mm, $R_{_{\square}}%
$=9.4M$\Omega$; (c) and (d) - InO$_{x}$, length=2.4mm, width=1.5mm,
$R_{_{\square}}$=20.1M$\Omega$. In these measurements, the conductance was
monitored under linear response conditions.\bigskip}%
\end{center}
\end{figure}

Note that, in response to thermal shocks, current-spikes are generated in
either case, and they are similar in shape to those produced `naturally' in
that they mimic the fast \textquotedblleft attack\textquotedblright\ and slow
recovery-tail form of the natural spikes (\textit{c.f.}, figure 10).%
\begin{figure}
[ptb]
\begin{center}
\includegraphics[
trim=0.000000in 1.408383in 0.000000in 0.576749in,
natheight=6.201600in,
natwidth=4.291200in,
height=3.333in,
width=3.3918in
]%
{wmf/Graph10.WMF}%
\caption{$\delta G(t)$ during natural (circles) and artificially (triangles)
produced spikes. The time-scale of the artificially produced spike is expanded
by 20 and its amplitude is shrunk by 20. Note that even though the artificial
spike has a much larger magnitude it is still much shorter than the natural
spike. Sample: InO$_{x}$, length=1mm, width=1.5mm, $R_{_{\square}}%
$=41.5M$\Omega$ (at the threshold drive).}%
\end{center}
\end{figure}
However, independent of the method by which they are produced, the artificial
spikes were much shorter than the `natural' ones typically are. In particular,
the recovery tail of the artificial spikes was always shorter than 3mS,
independent of the magnitude of the thermal shock or its sign. This 3mS is
presumably the time set by the thermal-inertia of the combination of the
substrate and sample-stage. By contrast, the duration of spikes produced by
above-threshold currents depend on sample parameters, most notably on its
carrier concentration $n$, and could be as large as 200mS. An example for the
dependence of the spike duration $\tau$ on n is illustrated in figure 11. Note
that for the sample with the largest $n$, $\tau$ exceeds the duration of the
artificial cooling-spikes by almost two orders of magnitude.%
\begin{figure}
[ptbptb]
\begin{center}
\includegraphics[
trim=0.000000in 1.408383in 0.000000in 0.576749in,
natheight=6.201600in,
natwidth=4.291200in,
height=3.333in,
width=3.3918in
]%
{wmf/Graph11.WMF}%
\caption{Spike shapes for three InO$_{x}$ samples with different carrier
concentration. Note the difference in the spike duration $\tau$.}%
\end{center}
\end{figure}

We shall argue below that the cooling events result from the formation of
(gaseous) cavities at the sample/helium interface. In this picture, the long
relaxation tail is associated with the re-liquefaction of the cavities. Before
addressing these issues in more detail however, we need to consider the
mechanism by which cavities are generated in the first place and in particular
the role of the sample parameters in triggering these cavities.

Led by the observation that spikes appear only when the system is in the
hopping regime (point 2 above), we review in the next section some of the
salient features of hopping conduction that distinguish it from diffusive
transport using the conventional percolation picture and taking into account
the discreteness of the hopping process. This treatment reveals the natural
reason for crossover behavior at a threshold current density; a feature
required by one of the empirical observations alluded to above. Moreover, the
dependence on the sample parameters agrees with the consequences of this
scenario, which also accounts for the exponential dependence of the spikes
frequency on current.

\section{IV -- Some relevant consequences of hopping conductivity}

The system we are dealing with, an Anderson insulator, is a degenerate Fermi
gas with spatial disorder sufficiently strong to cause localization of the
wave functions. Namely, the amplitude of the wave function is appreciable only
around a certain point in space $\mathbf{r}_{0}$ and decays as $\exp\left[
-\frac{|\mathbf{r}-\mathbf{r}_{0}|}{\xi}\right]  $ away from it. Charge
transport in such a system is controlled by the quantum mechanical transition
probability between localized states. The transition rate between sites i and
j is given by:

$\omega_{i,j}=\omega_{0}\exp\left[  -\frac{r_{ij}}{\xi}-\frac{\Delta E_{ij}%
}{k_{B}T}\right]  $\qquad(1)

Here $\omega_{0}$ is an attempt frequency (typically, $\omega_{0}%
\approxeq10^{12}-10^{13}$sec.$^{-1}$), $\xi$ is the localization radius,
$r_{ij}$ is the sites spatial separation, and $\Delta E_{ij}$ is their energy difference.

Since the system is disordered, $r_{ij}$ and $\Delta E_{ij}$ are random
variables distributed over some range. The exponential dependence of
$\omega_{i,j}$ on these variables (equation 1) leads to an extremely wide
distribution of transition rates. The macroscopic system may then be viewed as
a random-resistor-network in which each pair of sites $i,j$ is connected by a
Miller-Abrahams resistor \cite{19} $R_{ij}\varpropto\omega_{i,j}^{-1}$ The
wide distribution of the $R_{ij}$'s in the random-resistor-network leads to
several unique features of electronic transport in such a medium. The most
familiar feature is that the current in the system is carried by a
percolation-network \cite{20,21,22} involving a relatively small number of
localized sites while most of the sites that comprise the system are
disconnected from the current-carrying-network (CCN) and form pockets that are
effectively \textquotedblleft dead-wood\textquotedblright\ and some
dangling-off branches (\textquotedblleft dead-ends\textquotedblright) that do
not contribute to dc current. This is illustrated in figure 12.%
\begin{figure}
[ptb]
\begin{center}
\includegraphics[
trim=2.994057in 2.197072in 0.000000in 0.000000in,
natheight=4.797100in,
natwidth=6.299300in,
height=2.6394in,
width=3.3494in
]%
{wmf/Graph12.wmf}%
\caption{Schematic representation of the current carrying network in an
Anderson insulator. The lines are the current carrying paths. Stars represent
\textquotedblright bottleneck\textquotedblright\ resistors in the network.}%
\end{center}
\end{figure}
Each branch of the CCN is composed of a series of $R_{ij}$'s with different
values. The current through each branch is controlled by the largest $R_{ij}$
that percolates in the system (being inter-connected by smaller resistors).
This is called the critical-resistor $R_{C}$ or the \textquotedblleft
bottleneck\textquotedblright-resistor. The sheet resistance $R_{_{\square}}$
of a macroscopic two-dimensional sample is of the order of $R_{C}$. Upon a
change of temperature, electric field, magnetic field etc., the $R_{ij}$'s of
the system will in general change too, resulting in a modified CCN, with
different $R_{C}$ and $R_{_{\square}}$. The typical distance between $R_{C}$'s
on neighboring branches is called the percolation radius $L_{C}$ and it is the
measure of the CCN mesh-size.

The other feature that distinguishes hopping transport from that of metallic
conductivity is its discrete nature. The electron spends most of the time in a
localized state attempting to cross an effective barrier controlled by the
factor $\frac{r_{ij}}{\xi}+\frac{\Delta E_{ij}}{k_{B}T}$ (which is
typically$\gg$1) with an exponentially small probability while the actual
transition-time is very short. In other words, on a microscopic scale, charge
motion is a `stop-and-go' process marking the electron as a \textquotedblleft
bad driver\textquotedblright. A corollary of this picture is the emergence of
a \textquotedblleft critical current\textquotedblright\ $i_{c}$ related to the
transition rate through a bottleneck resistor. Using $e$ for the electronic
charge, $i_{c}$ may be expressed by:

$i_{c}=e\cdot\omega_{0}\cdot\exp\left[  -X_{C}\right]  $ where $X_{C}%
\equiv\left[  \frac{r_{ij}}{\xi}+\frac{\Delta E_{ij}}{k_{B}T}\right]  _{C}%
$\qquad\qquad(2)

To exceed $i_{c}$ an electron must be pushed towards site $i$ before the
electron already in this site has a chance to hop to site $j$. This situation
will be referred to as a jamming event.

To place an electron in j while the electron in i is still occupied entails an
energy barrier $\delta\varepsilon$ that has to be overcome. This
$\delta\varepsilon$ is of order of the associated Coulomb energy $e^{2}/\kappa
r_{ij}$ (where $\kappa$ is the dielectric constant of the medium). For
$r_{ij}\cong10-20\mathring{A}$, a typical inter-particle distance in these
films \cite{23} (also see: Vaknin \textit{et al} \cite{6}), and $\kappa\cong
$10, $\delta\varepsilon$ is $\cong100-50$meV respectively.

The system conductance will in general increase with the applied electric
field $F$ because there are always some $R_{C}$'s in the CCN for which $X_{C}$
decreases with $F$, specifically those where $\frac{r_{ij}}{\xi}\ll
\frac{\Delta E_{ij}}{k_{B}T}$. On the other hand, the CCN must also include
$R_{C}$'s for which $\frac{r_{ij}}{\xi}\gg\frac{\Delta E_{ij}}{k_{B}T}$, which
are insensitive to $F$. The latter group will become statistically more
important at larger fields as the system gets towards the `activationless'
regime \cite{24} which, as illustrated in figure 13, occurs near the region
where spikes first appear in the time window of the experiment.%
\begin{figure}
[ptb]
\begin{center}
\includegraphics[
trim=0.000000in 1.417956in 0.000000in 0.580300in,
natheight=6.246500in,
natwidth=4.356100in,
height=3.3347in,
width=3.4195in
]%
{wmf/Graph13.WMF}%
\caption{Resistance as a function of the applied bias for a typical In$_{2}%
$O$_{3-x}$ sample (length=1mm, width=1mm) Inset: Current as a function of the
applied bias plotted in accordance with equation 6 (dashed line). Arrows mark
the position of the threshold for spikes appearance.\bigskip Sample: In$_{2}%
$O$_{3-x}$, length=1mm, width=1mm, $R_{_{\square}}$=70M$\Omega$.\bigskip}%
\end{center}
\end{figure}
Therefore, it seems plausible that when the total current through the system
$I_{T}$ reaches a certain value, a jamming event might occur at some $R_{C}$.
In simple terms, when $F$ is increased, the rate by which electrons are forced
to cross the system is higher than the rate by which new conduction branches
are produced due to $F$. This is somewhat analogous to the traffic-congestion
problem in urban areas; with time, the number of cars on the road usually
increases at a larger rate than the rate of the production of new roads,
leading to the annoying traffic-jam during rush hours.

The total current at which the likelihood of a jam event becomes appreciable
can be estimated as follows. Assume a two-dimensional system, namely, a film
of thickness \cite{25} $d\ll L_{C}$, and width $W$. The critical current
$i_{c}$ through a typical branch will then be reached when $I_{T}\approx
i_{c}\cdot W/L_{C}$. Using equation 2 and noting that $R_{_{\square}}\approx
R_{C}\approx R_{0}\cdot\exp\left[  X_{C}\right]  $ this happens when the (2D)
current density $J_{C}^{S}$ is $\approx\frac{e\omega_{0}R_{0}}{R_{_{\square}%
}L_{C}}$. It is often found empirically \cite{26,27} in two-dimensional
hopping systems that $R_{0}$ is of the order of the quantum-resistance
$\approx\frac{h}{e^{2}}$. Using this relation one gets:

$J_{C}^{S}=\frac{I_{T}}{W}\approx\frac{h\omega_{0}}{eL_{C}R_{_{\square}}}%
$\qquad(3)

A jamming event is then likely to occur once the \textit{current density}
through the system exceeds a value, which is inversely proportional to the
sheet-resistance $R_{_{\square}}$. Alternatively, the threshold condition can
be expressed as a critical voltage drop $V_{c}$ across a bottleneck resistor:

$V_{c}=e\omega_{0}R_{0}\approx\frac{\omega_{0}h}{e}$\qquad(4)

We now show that the onset for the conductance-spikes in our samples, as well
as their frequency as function of the local voltage (or the global
current-density), follow the behavior expected by the above considerations.
First, consider the correlation between the threshold-current density
$J_{C}^{S}$ at which the spikes appear and the sample $R_{_{\square}}$ shown
in figure 14.%
\begin{figure}
[ptb]
\begin{center}
\includegraphics[
trim=0.000000in 1.288653in 0.000000in 0.580924in,
natheight=6.246500in,
natwidth=4.356100in,
height=3.4359in,
width=3.4195in
]%
{wmf/Graph14.WMF}%
\caption{The 2D current density at threshold as a function of sample sheet
resistance (evaluated at threshold). The open squares and solid circles are
InO$_{x}$ and In$_{2}$O$_{3-x}$ samples, respectively.\bigskip}%
\end{center}
\end{figure}
This figure summarizes results of more than 40 In$_{2}$O$_{3-x}$ and InO$_{x}$
samples with vastly different geometries; in particular, it includes samples
with width $W$ in the range 20$\mu$m to 3cm. This three orders of magnitude
spread should be compared with a mere factor of $\pm$4 scatter in the data in
the figure. Thus, the correlation between $J_{C}^{S}$ and $R_{_{\square}}$ is
quite suggestive. The dashed line in the figure is calculated by equation 3
using $\omega_{0}=6\cdot10^{12}\sec^{-1}$ and $L_{C}$=1$\mu$m, both are
reasonable values for a hopping system. The use of a disorder-independent
percolation radius needs justification. For a Mott VRH in 2D (which is the
case for the In$_{2}$O$_{3-x}$ samples) one expects $L_{C}$ to scale as
$\left[  r\cdot\left(  \frac{T_{0}}{T}\right)  ^{\frac{1}{3}}\right]
^{\frac{1}{\nu}}$ where $\nu$ is of order 1 \cite{28}. This would then lead to
$L_{C}$ going like $\xi\cdot\left(  \frac{T_{0}}{T}\right)  ^{\frac{2}{3}}$
and combined with $T_{0}\propto\xi^{-2}$, and at given temperature,
$L_{C}\propto\xi^{-\frac{1}{3}}$. For the range of $R_{_{\square}}$ studied
here (10M$\Omega$-1G$\Omega$), the associated $\xi^{\prime}s$ range from
$20\mathring{A}$ to $5\mathring{A}$ respectively, which means that the
variation of $L_{C}$ is a mere factor of 1.6, which is smaller than the
scatter in the data in figure 14. Even this factor is compressed by the
voltage; Note (equation 4) that the threshold is characterized by a
\textit{constant} local voltage. Since $L_{C}$ is reduced by voltage, the more
so the larger `equilibrium $L_{C}$' is, the effective range of $L_{C}$ is
smaller than the estimate based on near-equilibrium conditions. It seems
harder to understand why the results for the InO$_{x}$ samples should fall on
the same line as those of the crystalline samples. As mentioned above (section
II), InO$_{x}$ samples (especially those with high $n$) show a different
hopping law than In$_{2}$O$_{3-x}$ samples of similar $R_{_{\square}}$. There
is thus no a-priori reason to assume that In$_{2}$O$_{3-x}$ and InO$_{x}$
samples of equal resistance should have the same $L_{C}$. Nevertheless,
independent estimates of $L_{C}$ in these systems appear to be in the same
range of values. For In$_{2}$O$_{3-x}$ films in this range of disorder,
$L_{C}$ values ranging between 0.3 to 0.5$\mu$m were found based on the
magnitude of the conductance fluctuations as a function of sample size
\cite{11}. In high carrier concentration InO$_{x}$ films, a value of 0.3$\mu$m
$L_{C}$ may be estimated by extrapolating the results of Frydman et al
\cite{29}. The latter are based on measuring the resistance versus thickness.
Interestingly, a similar value for $L_{C}$ of 1$\mu$m was estimated based on
the magnitude of mesoscopic fluctuations for insulating films of granular
nickel \cite{30}, a system that differs markedly from both, crystalline and
amorphous indium-oxide.

The local voltages $V_{c}$ associated with the data points in figure 14 can be
estimated using the expression: $\frac{I_{T}^{c}R_{_{\square}}L_{C}}{W}$ and
they vary in the range 12-70mV while the sheet resistance $R_{_{\square}}$ (at
threshold) changes by more than two orders of magnitude. This range for
$V_{c}$ compares favorably with equation 4 that for $\omega_{0}=6\cdot
10^{12}\sec^{-1}$ yields $V_{c}$=25mV.

Furthermore, the same $V_{c}$ seems to be in control of the dependence of the
spikes (average) frequency $\omega$ on the current near threshold shown above
(figures 4 and 5). The extremely fast $\omega(I_{T})$ is suggestive of an
activated process, which turns out to be of the Arrhenius type. This is
illustrated in figure 15 for several \cite{31} of the studied samples where we
plot $\omega$ as function of $V=\frac{I_{T}R_{_{\square}}L_{C}}{W}$
(normalized by the respective $V_{c}$ for each sample). It is tempting to
analyze these results using a relation of the form:

\qquad$\omega=\omega_{0}\exp\left[  -\frac{e(V_{c}-V)}{k_{B}T^{\ast}}\right]
$\qquad(5)\newline with the rationale that the local voltage $V$ increases the
probability to cause a jam by reducing the barrier $V_{c}$. Note that the
$T^{\ast}$ used in equation 5 is unlikely to be the same as the bath
temperature $T$ in the present case. Recall that we are dealing here with a
far from equilibrium situation and there is no unique way to estimate an
effective temperature $T^{\ast}$. As remarked above, the spikes occur when the
field $F$ brings the system near (but somewhat below) the activationless
conductance regime where the `temperature' of the electrons is controlled by
\cite{24} $F$. For the lack of anything better, it seems plausible to estimate
$T^{\ast}$ from the resistance at the threshold field by comparing it to the
\textquotedblleft ohmic\textquotedblright\ $R(T)$ data (namely, using the
resistance of the sample as a `thermometer'). For the samples used in figure
15 we get $k_{B}T^{\ast}$ in the range 0.5meV to 0.9meV and using these data
with $\omega_{0}=6\cdot10^{12}\sec^{-1}$ in equation 5, gives a range for
$V_{c}$ of 25-100mV. Note that these values of $V_{c}$ overlap with the range
of energies associated with $\delta\varepsilon=\frac{e^{2}}{\kappa r}$
discussed above.%
\begin{figure}
[ptb]
\begin{center}
\includegraphics[
trim=0.000000in 1.417956in 0.000000in 0.580300in,
natheight=6.246500in,
natwidth=4.356100in,
height=3.3347in,
width=3.4195in
]%
{wmf/Graph15.WMF}%
\caption{Frequency of the spikes as function of the voltage across a
bottleneck resistor (see text). Different symbols are used for different
samples.\bigskip}%
\end{center}
\end{figure}

There are then several independent indications that the jam-scenario may be
involved in controlling the onset as well as the frequency of appearance of
the conductance spikes.

The question is what is then the role of the He bath: Why the electronic
mechanism does not generate spikes unless the sample is in contact with liquid
helium. The answer is -- it presumably does. However, the electronic processes
that take place due to a traffic-jam should usually die out very quickly and
escape observation. Consider that a jamming event has occurred at some point
in the sample. The population of electronic states near a jammed site may then
look as in figure 16.%
\begin{figure}
[ptb]
\begin{center}
\includegraphics[
trim=2.853583in 2.074249in 0.000000in 1.815989in,
natheight=6.283700in,
natwidth=6.299300in,
height=1.8801in,
width=2.6939in
]%
{wmf/Graph16.wmf}%
\caption{A schematic description of the localized states in the CCN near a
bottleneck resistor (that for $i\ll i_{C}$ is assumed to be i,j), for small
currents ($i\ll i_{C}$ -- lower configuration) and for a typical jam event
($i\geq i_{C}$ -- upper configuration). Light circles are unoccupied sites and
dark circles are occupied sites. $\delta\varepsilon$ is symbolized here as the
sum of the displacements of the sites k and i from the baseline.\bigskip}%
\end{center}
\end{figure}
Note that, for $i\geq i_{c}$ an effective energy barrier $\delta\varepsilon$
is created due to the repulsion between the electrons occupying sites k and i.
Such a configuration is a high-energy state and it will quickly dissipate
itself, say, by the electron in $i$ hopping away to another site. The
transition time associated with a single-particle hop is quite short. This may
be estimated from $\omega^{-1}\approx\frac{\omega_{0}^{-1}R_{_{\square}}e^{2}%
}{h}$ to be $\approx10^{-9}\sec$ for $R_{_{\square}}\simeq$100M$\Omega$.
Therefore, the duration of an electronic-jam ought to be much shorter than the
spike widths we typically observe (\textit{c.f.}, figure 11). The latter
therefore must involve another agent such as the one suggested next.

\section{V - A mechanism for cooling events triggered by the electronic
traffic-jams}

We suggest that the spikes are the response of the sample to cooling events
associated with nucleation of He cavities at the sample/liquid interface.
These are triggered by the electronic jams, which thus control the onset and
frequency of the events, while the duration of a spike is determined by the
lifetime of the formed cavities. Gaseous helium cavities may be produced by
heat-pulses \cite{32} or by acoustic-wave-bursts \cite{33,34}, and in
particular, the threshold for a cavity formation on glass surfaces
(heterogeneous cavitation) is quite low \cite{35}. Both heating-pulses and
acoustic-waves are potential products of energy-bursts involved in the jam
events. That a brief heat-pulse may emanate from a jam event seems obvious.
Emission of acoustic wave needs more elaboration. To see how this might happen
note that when a jam is \textquotedblleft on\textquotedblright\ an electric
field appears for a brief moment across a bottleneck resistor. This field is
of the order of 10$^{5}$V/cm, which is quite large. While this field is on,
the medium will polarize to some degree. This in turn will be accompanied by a
mechanical deformation (electrostriction). When the jam goes \textquotedblleft
off\textquotedblright\ the stress on the medium is relieved, and this cycle of
`pull-and-let-go' is very likely to generate acoustic waves much like plucking
a string does \cite{36}. This effect is possible since the "on" time of the
jam is longer than $\omega_{D}^{-1}$ ($\omega_{D}$ is the Debye frequency of
the medium which is of the order of 10$^{12}$-10$^{13}\sec^{-1}.$) even for
R$_{C}$=1M$\Omega$. It is hard to tell which of these two agents is more
dominant in our experiments and perhaps they are complementary. In fact, our
scenario is somewhat analogous to the way Glaser's bubble chamber detects
high-energy particles. The operation of the bubble chamber involves
super-heating the liquid followed by a sudden drop of pressure to make visible
the trajectory of a fast moving particle. Super-heating in our picture is
facilitated by the accentuated local heating inherent to the inhomogeneous
mode of hopping transport. The local temperature near a bottleneck resistor
(on scale of a thermal phonon wavelength $\approx$1000$\mathring{A}$) could
conceivably be a fraction of a degree K above the temperature of the
non-conducting part of the sample. The effective barrier for cavitation will
be lower at these hot spots and bubble nucleation will most likely start at
one of these spots. Below the $\lambda-$ point, such hot spots will be
eliminated by the superfluid counter-flow and this presumably is the main
reason for the disappearance of the spikes (\textit{c.f.}, figure 8).

Once a cavity nucleates and starts to expand, it sucks energy from the medium
(the latent heat of transforming liquid into gas) thus producing local cooling
\cite{32}. This is detected as a drop of the conductance -- the fast
\textquotedblleft attack\textquotedblright\ of the spike. The trailing edge of
the spike is controlled by the re-condensation of the gaseous cavity. Hence,
the duration $\tau$ of a spike depends on the volume of the cavity. A simple
estimate \cite{37} for re-condensation of a bubble with diameter $D$ in cm
yields $\tau\approx$10$^{8}D^{3}\sec$. To account for the spike duration-times
observed in figure 11 (varying in the range 1-200mS) the respective cavity
diameters has to be in the range of 2-10$\mu$m. Even the largest bubbles in
this range are too small to be optically imaged by our current techniques yet
they posses large enough thermal-mass to sustain the cooling effect for a long time.

Further support for the relevance of cavitation is the dependence of the spike
frequency on the vapor pressure of the helium bath. As illustrated in figure
17a, the average spike frequency $\omega$ decreases with pressure $P$. This is
consistent with the assumption that cavities are facilitated by the negative
pressure of an acoustic wave. To keep $\omega$ constant when $P$ increases
$I_{T}$ has to be slightly (few percents) increased. However, under these
conditions, the magnitude of the spikes decreases extremely fast with $P$ as
shown in figure 17b. Over the range of $\Delta P$ shown in the figure, the
sample conductance changes by a negligible amount, and by applying artificial
cooling-bursts in this region it was found that the sample's sensitivity as a
bolometer is independent of $P$. It must therefore be concluded that the
decrease of $\Delta G/G$ with pressure results from the diminished cooling
power of the cavities. We interpret this decrease as indication that the
phenomenon occurs near the region where the heat of vaporization changes very
rapidly with $P$ (and $T$) as it would near the critical temperature.%
\begin{figure}
[ptb]
\begin{center}
\includegraphics[
trim=0.000000in 0.773317in 0.000000in 0.452246in,
natheight=6.246500in,
natwidth=4.356100in,
height=3.9366in,
width=3.4195in
]%
{wmf/Graph17.WMF}%
\caption{(a) The frequency of the spikes as a function of change of the vapor
pressure of the helium bath. (b) The relative magnitude of the spikes as a
function of the vapor pressure of the helium bath. This experiment was done by
pressurizing the storage dewar, and monitoring the results 'on the fly'. The
entire experiment took about 7 minutes to accomplish, during which the
temperature rise was negligible as indicated by the change in G. $\Delta$P=0
is related to the atmospheric pressure (693Torr). Sample: InO$_{x}$,
length=0.7mm, width=1mm, $R_{_{\square}}$=1.7M$\Omega$ (at the threshold
drive).\bigskip}%
\end{center}
\end{figure}

We have considered another scenario for cooling, based on the notion that
bubbles, produced by local Joule-heating, grow and eventually float up thereby
the sample is re-cooled by helium counter flow. There are a number of problems
with such a scenario. To leave the surface of the sample a gaseous bubble has
to be rather big, of the order of few hundred microns, and the energy required
to produce it at the rate observed is incompatible with the input Joule-energy
and clearly inconsistent with the exponential dependence of this rate. In
addition, we made a number of tests to explore the possibility that bubbles
may form to be big enough do leave the sample. Two of these are described below.

In the first experiment, two samples were lowered into the helium bath and
were attached to the probe parallel to one another and to the floor. The lower
one was biased with current so that it produced spikes. The upper sample
(3-4mm above the \textquotedblleft active\textquotedblright\ one) was biased
with sub-threshold current and showed no hint of any thermal shock that ought
to have been recorded had a gaseous bubble emanating from the lower sample hit
its surface.

In another experiment, an electromechanical \textquotedblleft
woodpecker\textquotedblright\ was employed with the idea that mechanical
shocks may dislodge some \textquotedblleft ripe\textquotedblright\ bubble and
thus synchronize spike appearance. The device used was a small loudspeaker
cone to which a small plastic beak was glued and the rig was attached on top
of the probe such that it could periodically transmit a mechanical kick down
to the sample. The latter was biased with near threshold current that
naturally produced occasional spikes. Various combinations of the
\textquotedblleft woodpecker\textquotedblright\ frequency and amplitude were
tried as well as several choices of bias current. No correlation was found
between the mechanical shocks and the spikes appearance. The negative results
of these experiments give further reasons to rule out the cooling mechanism
based on \textquotedblleft bubble-emission\textquotedblright.

\section{VI - Further results and discussion}

To shed more light on the cooling events and the way they are produced we
performed the following experiments. In the first, the idea was to further
check on the thermal nature of the effect by using a sample (biased with a
sub-threshold current) as a bolometer to detect the pulse that, by assumption
emanates from the `active' sample. For that purpose, two samples were
deposited on the same substrate as two cross strips separated by a thick
insulating barrier (5000$\mathring{A}$~of SiO$_{2}$). As shown in figure 18
the `bolometer' records a signal that is synchronized with and mimics the main
features of the spike observed in the $G(t)$ of the active sample. The
magnitude of the signal at the detector sample is considerably reduced (note
that the sensitivity of the `bolometer' given by $\frac{dG}{dT}$ is greater
than that of the active sample because it is used with a smaller bias
current). In addition, the spike appears somewhat delayed and broader relative
to the original, which is consistent with a diffusive propagation process.%
\begin{figure}
[ptb]
\begin{center}
\includegraphics[
trim=0.000000in 1.417956in 0.000000in 0.580300in,
natheight=6.246500in,
natwidth=4.356100in,
height=3.3347in,
width=3.4195in
]%
{wmf/Graph18.WMF}%
\caption{$\delta G(t)$ generated by the active sample (open squares) and
recorded by the detecting sample (solid circles) (see text for details).
Samples: generator - InO$_{x}$, length=2.4mm, width=0.7mm, $R_{_{\square}}%
$=0.8M$\Omega$ (at the threshold drive); detector - InO$_{x}$, length=1.5mm,
width=1mm, $R_{_{\square}}$=6.7G$\Omega$.\bigskip}%
\end{center}
\end{figure}

While the appearance of the spikes is not reflected in the (averaged) $R$ vs.
$F$ plots (\textit{c.f}., figures 6 \& 13), it is readily observed in
derivative measurements. The advantage of this type of measurement is that it
makes it possible to check for hysteresis in the threshold current by changing
the sweep rate. We studied the differential conductance $\frac{dI}{dV}$ of
some samples as a function of the bias voltage V focusing on a limited range
straddling the threshold for spikes appearance. A typical $\frac{dI}{dV}(V)$
plot is shown for a In$_{2}$O$_{3-x}$ sample in figure 19.%
\begin{figure}
[ptb]
\begin{center}
\includegraphics[
trim=0.000000in 1.417956in 0.000000in 0.580300in,
natheight=6.246500in,
natwidth=4.356100in,
height=3.3347in,
width=3.4195in
]%
{wmf/Graph19.WMF}%
\caption{Dynamic conductance (dI/dV) versus voltage over a region including
the threshold for spikes appearance. Open and solid triangles represent data
taken upon sweeping the applied bias (V) up and down respectively. Sweep rate
is 0.08V/sec. Note the small hysteresis. The dashed line shows the
differential conductance expected by equation 7. Note that the experimental
dI/dV is lower than the theoretical value even at voltages much larger than
the voltage where individual spikes cannot be resolved in the time domain.
This sample is the same as in figure 13.}%
\end{center}
\end{figure}
Note that at $V\approx V^{\ast}$, where $V^{\ast}$ is the threshold value for
spike appearance, $\frac{dI}{dV}$ exhibits a sharp change and above $V^{\ast}$
the conductance is smaller than the theoretical curve obtained as follows. As
shown in figure 13, the spikes in this sample appear in the voltage regime
somewhat below the transition to activationless hopping. The current-voltage
characteristics in 2D take the form \cite{38}:

$\log[I]\varpropto-\left(  \frac{V_{0}}{V}\right)  ^{\frac{1}{3}}$\qquad
\qquad(6)\newline from which the differential conductance is obtained as:

$\frac{dI}{dV}\varpropto\exp[-\left(  \frac{V_{0}}{V}\right)  ^{\frac{1}{3}%
}]\cdot V^{-\frac{4}{3}}$\qquad(7)\newline this is plotted in figure 19 with
the value of $V_{0}$ extracted by fitting equation 6 to the data in figure 13.
The difference between this theoretical curve and the experimental $\frac
{dI}{dV}$ is 2.1-3.4\% while the maximum amplitude of the spikes for this
sample is 0.02\%. In other samples, we noticed that $\frac{dI}{dV}$ above
threshold increase at a slower rate than the natural trend observed below
$V^{\ast}$ leading in some samples to even bigger discrepancies \cite{39}.
Therefore, the diminished conductance above threshold cannot be simply
explained by the fact that the spikes (that are associated with negative $dI$)
\textquotedblleft pull-down\textquotedblright\ the signal. The reason for this
discrepancy might be due to the way the cavitation processes, that are fed by
the applied field, interfere with the photon-assisted-hopping processes
involving the same field \cite{40}. This problem is currently under further investigation.

As a further test of the conjecture that the spike phenomenon involves hopping
transport, we made a preliminary study of the effect of a magnetic field on
the spikes. For this purpose, we chose to work with the crystalline version of
indium-oxide that has been extensively studied \cite{41,42} and where the
magneto-resistance (MR) is well understood as being due to orbital and spin
effects \cite{42} The sign of the MR is negative or positive when the dominant
mechanism is quantum-interference effect (orbital) or spin-alignment
(isotropic) respectively. The relative contribution of the orbital part can be
controlled by changing the angle between the sample plane and the field
direction. It is thus possible to compare the results at high field $H_{1}$
with the results at small field $H_{2}$ while adjusting things such that
$R_{_{\square}}(H_{1})=R_{_{\square}}(H_{2})$. Note that while the sample is
the same from the point of view of the measurement circuit, it must be
microscopically different; both configurations have similar bottleneck
resistors $R_{C}$'s but at least some of these are placed at different
locations relative to the sample axes. In other words, the current carrying
network must be different. To see why that must be so it is enough to note
that the spin-alignment mechanism eliminates from the CCN doubly occupied
states, a process that is complete for $g\mu H>k_{B}T$ \cite{43}. The result
of this magnetic field stratagem is illustrated in figure 20 for one of the
two samples studied so far.%
\begin{figure}
[ptb]
\begin{center}
\includegraphics[
trim=0.000000in 1.417956in 0.000000in 0.580300in,
natheight=6.246500in,
natwidth=4.356100in,
height=3.3347in,
width=3.4195in
]%
{wmf/Graph20.WMF}%
\caption{Time traces at two different magnetic field for In$_{2}$O$_{3-x}$
sample (length=1mm, width=2mm, $R_{_{\square}}$=7M$\Omega$ at the threshold
drive). Note the different spike characteristics observed. Inset:
Magneto-conductance as a function of magnetic field for the sample.\bigskip}%
\end{center}
\end{figure}
In this particular case, the 9T field has presumably shifted the
nucleation-site of the cavity resulting in a different looking spike. This is
a plausible scenario; the hottest spot on the sample would be the preferred
site for cavity nucleation and this spot is probably at or near one of the
critical resistors in the CCN. Therefore, when the CCN is changed by the field
the spike might originate at a different location. And exhibit somewhat
different shape much like when using another sample from the same batch.

As noted in section III, in quite a few cases, the spikes near threshold seem
to have an identical shape. This \textquotedblleft
mesoscopic\textquotedblright\ behavior in macroscopic samples is intriguing;
Both, $L_{C}$ and the cavities diameter, which are the relevant parameters,
are much smaller than the size of the sample (which is typically 1mm across).
Several factors may be involved in bringing about this behavior. In our model,
it was implicitly assumed that the bottleneck resistors in the CNN are all
equal. This may be a reasonable assumption for assessing the threshold current
but the $R_{C}$'s are probably distributed over a considerable range,
especially when the disorder is large. The jam events near threshold
presumably involve the $R_{C}$'s at the far tail of this distribution. The
question is how many `active' $R_{C}$, are there over the $\approx$3\% range
of current, where individual spikes are resolved in 15-20\% of the samples.
Alternatively, there may be a `preferred' site for cavity nucleation, which is
triggered by any of a number of (simultaneous) jam events. In other words, we
see no compelling reason that the cavity must be nucleated at the generating
$R_{C}$ site though probably not too far from it.

An important issue that needs further elucidation is what determines the spike
duration in a given sample. The spike duration $\tau$ and the relative
amplitude of the spikes $\delta G/G$ are usually correlated as illustrated in
figure 21.%
\begin{figure}
[ptb]
\begin{center}
\includegraphics[
trim=0.000000in 1.288653in 0.000000in 0.580924in,
natheight=6.246500in,
natwidth=4.356100in,
height=3.4359in,
width=3.4195in
]%
{wmf/Graph21.WMF}%
\caption{Upper graph - Spikes duration (defined in figure 11) as a function of
carrier concentration. Lower graph -- relative magnitude of the spikes for the
same series of InO$_{x}$ samples. The dashed curves are guides for the
eye.\bigskip}%
\end{center}
\end{figure}
These results were obtained using InO$_{x}$ samples where n could be changed
over a large range by varying the In/O ratio. To allow comparison, all samples
in this figure have similar dimensions of 1x1mm. No correlation between $\tau$
and disorder of a given sample was found. Note that longer $\tau$'s are
statistically accompanied by bigger $\delta G/G$, and both decrease rapidly
below a carrier-concentration of $\approx6\cdot10^{19}$cm$^{-3}$. The
correlation between $\tau$ and $\delta G/G$ is natural in our scenario; larger
cavities presumably produce more cooling and last longer \cite{44}. The
dependence of both on n however is more intriguing in that it suggests that
the phenomenon may disappear in the limit of small carrier concentration.
Indeed, in samples with $n$%
$<$%
5$\cdot$10$^{19}$cm$^{-3}$ the spikes are so short and have so small relative
magnitude as to make their detection very difficult as compared with the
prominent spikes produced by samples with larger $n$ (\textit{c.f.}, figure
11). A plausible way to understand this trend is to note that samples with
larger $n$ are more likely to produce cavities due to their stronger
electron-electron interactions; An Anderson insulator lacks metallic
screening, which in turn means that the inter-particle interaction is stronger
in a system with higher carrier concentration \cite{6}. Hence, such a system
has more \textquotedblleft kick\textquotedblright\ to produce a bigger cavity
by the mechanism described above.

This mechanism for spike production is generic; it should apply to any system
provided that two conditions are met; a) transport is by hopping, and b)
interactions are strong. Therefore, natural candidates for such behavior are
electron-glasses that are glassy by virtue of the very same two ingredients
\cite{5}. Led by this consideration, we tested a granular aluminum sample, a
hopping system that was recently shown to exhibit electron-glass features
\cite{45} The results of such an experiment are shown in figure 22
demonstrating conductance spikes and bias dependence that are similar to the
respective behavior in crystalline and amorphous indium-oxide films. It would
be of interest to test our picture in other hopping systems whether or not
they exhibit glassy features, and in other dimensionalities as well.%
\begin{figure}
[ptb]
\begin{center}
\includegraphics[
trim=0.000000in 1.417956in 0.000000in 0.580300in,
natheight=6.246500in,
natwidth=4.356100in,
height=3.3347in,
width=3.4195in
]%
{wmf/Graph22.WMF}%
\caption{$\delta G(t)$ for a granular Al sample measured at different currents
near the threshold (the current values, in $\mu$A, are indicated for each
trace). Sample: length=1mm, width=1mm, thickness=200$\mathring{A}$,
$R_{_{\square}}$=33.7M$\Omega$ (at the threshold drive).}%
\end{center}
\end{figure}

Finally, we wish to comment on the nature of the transition that is
characterized by the appearance of the conductance spikes. In some respects,
the phenomenon resembles a phase transition. The spikes appear and disappear
rather suddenly over an extremely narrow range of current, and in several
cases, hysteretic behavior was observed. It should however be noted that the
underlying picture of traffic jam is essentially an extremely sharp crossover
rather than a phase transition. The fact that a threshold current is assigned
to a given sample is merely a result of the exponential dependence expressed
by equation 5 and the finite time-window of the observation. The hysteretic
behavior is probably related to the formation of cavities, a process that is
more likely to be a real (non-equilibrium) phase transition than what has been
termed here a traffic jam.

A relevant question is the role of many-particle transitions that are expected
\cite{46} to be important in a hopping system with strong electron-electron
interactions. This aspect of hopping transport was omitted from our treatment
on the assumption that many particle transitions become less important in the
non-ohmic regime. Obviously, transport by many particles, correlated
transitions could obviate the traffic jam events that are depicted here as
resulting from single-particle transitions. In principle, extending the study
of the dependence of the spike frequency on drive current to the low frequency
regime might shed some light on this question.

In summary, we presented data on a new kind of conductance noise characterized
by the appearance of downward-going spikes in the conductance versus time
traces. This was shown to be a non-equilibrium phenomenon peculiar to the
hopping regime. A model taking into account the specific features of hopping
transport is shown to be consistent with several aspects of the phenomenon. In
particular, the model explains the observation that the onset for the spike
appearance and their average frequency is controlled by the current density.
It also predicts certain dependencies on sample parameters in agreement with
experiments on many different samples. The purely electronic consequences of
this electronic-jam picture are expected to occur at microwave frequencies and
would be difficult to observe. We ascribe the spikes to cooling events
produced by cavitation phenomena. In our scenario, cavities are nucleated in
response to pulses of energy (heat and/or acoustic waves) generated and
synchronized by the electronic jams events. The time scales associated with
the re-condensation of the cavities is long enough to facilitate observation
of individual events in much the same vein that bubble chambers are used in
tracking fast moving elementary particles.

From the point of view of cavitation phenomena it is worth noting the
similarity of our picture and the work of Sinha et al \cite{32}. Sinha et al
used in their experiment electrical pulses fed into a Bi crystal to generate
cavities in the helium. The Bi sample doubled as a bolometer detecting the
thermal effects that resulted from cavitation, just as our samples are
presumed to do. Their process differs from ours mainly in that in the hopping
system electrical pulses are generated naturally by the sample, which
essentially acts like a current-controlled pulse-generator. This inherent
property of the hopping regime should make these systems attractive for the
study of cavitation phenomena.

We acknowledge illuminating discussion with H. Maris, M. Pollak, and O. Agam.
This research has been supported by the Binational US-Israel Science
Foundation and by The Israeli Academy for Sciences and Humanities.


\begin{thebibliography}{99}
\bibitem {17}Identification of the threshold is much easier using the spectrum
as it retains its anomalous shape even for drive current much larger than the threshold.
\bibitem {25}The films used here are 2D even by the more stringent criterion
that the hopping length is larger than the thickness.
\bibitem {31}This trend could be reliably studied only in samples that exhibit
prominent spikes and where individual spikes could be identified.
\bibitem {37}H. Maris, (Private communication). The picture he suggested is
based on that in order for the bubble to collapse, the vapor in it must
condense and the heat that is liberated needs to be conducted away. The rate
at which this happens is determined by the volume of the cavity (in addition
to the heat conductivity of the liquid), which in turn determines the recovery time.
\bibitem {39}Comparing the experimental dI/dV to equation 7 is for convenience
only. With that we actually underestimate the discrepancy because this
activationless expression has a weaker dependence on V than the realistic
dI/dV for V smaller than the activationless threshold.
\bibitem {44}Note that the temperature coefficient of the resistance also
plays a role in determining the magnitude of the spike.
\end{thebibliography}
\end{document}